\documentclass[12pt]{article}
\usepackage{amsmath}
\usepackage{latexsym}
\usepackage{bbm}
\usepackage{color}
\usepackage{graphicx}

\def \AP {{\it Annals Phys.}, }

\def \ATMP {{\it Adv. Theor. Math. Phys.}, }

\def \NC {{\it Nuovo Cim.}, }
\def \NCL {{\it Nuovo Cim. Lett.}, }
\def \NP {{\it Nucl. Phys.}, }
\def \NPPS {{\it Nucl. Phys. Proc. Suppl.}, }
\def \PL {{\it Phys. Lett.}, }
\def \PR {{\it Phys. Rev.}, }

\def \PRL {{\it Phys. Rev. Lett.}, }
\def \PTP {{\it Prog. Theor. Phys.}, }

\def\a{\alpha}

\def\b{\beta}
\def\g{\gamma}

\def\d{\delta}

\def\e{\varepsilon}

\def\m{\mu}

\def\r{\rho}

\def\s{\sigma}
\def\t{\tau}

\def\G{\Gamma}

\def\pa{\partial}

\def\half{\frac{1}{2}}

\def\IC{\mathbbm C}

\def\IR{\mathbbm R}
 
\def\IZ{\mathbbm Z}

\def\half{\frac{1}{2}}

\def\and{{\rm and}}

\def\ie{{\it i.e.,} }

\begin{document}
\vspace*{-.6in} \thispagestyle{empty}
\begin{flushright}
CALT-68-2657
\end{flushright}
\baselineskip = 18pt

\vspace{1.5in} {\Large
\begin{center}
THE EARLY YEARS OF STRING THEORY:\\A PERSONAL PERSPECTIVE
\end{center}} \vspace{.5in}

\begin{center}
John H. Schwarz
\\
\emph{California Institute of Technology\\ Pasadena, CA  91125, USA}
\end{center}
\vspace{1in}

\begin{center}
\textbf{Abstract}
\end{center}
\begin{quotation}
\noindent This article surveys some of the highlights in the
development of string theory through the first superstring
revolution in 1984. The emphasis is on topics in which the author
was involved, especially the observation that critical string theories
provide consistent quantum theories of gravity and the proposal to use
string theory to construct a unified theory of all fundamental
particles and forces.
\end{quotation}

\vspace{1in}
\centerline{\it Based on a lecture presented on June
20, 2007 at the Galileo Galilei Institute}

\newpage

\pagenumbering{arabic}

\section{Introduction}

I am happy to have this opportunity to reminisce about the origins
and development of string theory from 1962 (when I entered graduate
school) through the first superstring revolution in 1984. Some of
the topics were discussed previously in three papers that were
written for various special events in 2000
\cite{Schwarz:2000yx,Schwarz:2000yy,Schwarz:2000dm}. Also, some of
this material was reviewed in the 1985 reprint volumes
\cite{Schwarz:1985ws}, as well as the string theory textbooks
\cite{Green:1987cup,Becker:2007cup}. In presenting my experiences
and impressions of this period, it is inevitable that my own
contributions are emphasized.

Some of the other early contributors to string theory have presented
their recollections at the Galileo Galilei Institute meeting on
``The Birth of String Theory'' in May 2007. Since I was unable to
attend that meeting, my talk was given at the GGI one month later.
Taken together, the papers in this collection should convey a fairly
accurate account of the origins of this remarkable
subject.\footnote{Since the history of science community has shown
little interest in string theory, it is important to get this
material on the record. There have been popular books about string
theory and related topics, which serve a useful purpose, but there
remains a need for a more scholarly study of the origins and history
of string theory.}

The remainder of this paper is divided into the following sections:

\noindent $\bullet$ 1960 -- 68: The analytic S matrix (Ademollo,
Veneziano)

\noindent $\bullet$ 1968 -- 70: The dual resonance model (Veneziano,
Di Vecchia, Fairlie, Neveu)

\noindent $\bullet$ 1971 -- 73: The RNS model (Ramond, Neveu)

\noindent $\bullet$ 1974 -- 75: Gravity and unification

\noindent $\bullet$ 1975 -- 79: Supersymmetry and supergravity
(Gliozzi)

\noindent $\bullet$ 1979 -- 84: Superstrings and anomalies (Green)

\noindent For each section, the relevant speakers at the May meeting
are listed above. Since their talks were more focussed than mine,
they were able to provide more detail. In one section (gravity and
unification) my presentation provided more detail than the others.

\section{1960 -- 68: The analytic S matrix}

In the early 1960s there existed a successful quantum theory of the
electromagnetic force (QED), which was completed in the late 1940s,
but the theories of the weak and strong nuclear forces were not yet
known. In UC Berkeley, where I was a graduate student during the
period 1962 -- 66, the emphasis was on developing a theory of the
strong nuclear force.

I felt that UC Berkeley was the center of the Universe for high
energy theory at the time. Geoffrey Chew (my thesis advisor) and
Stanley Mandelstam were highly influential leaders. Also, Steve
Weinberg and Shelly Glashow were impressive younger faculty members.
David Gross was a contemporaneous Chew student with whom I shared an
office.\footnote{It was a particularly nice office, which was being
reserved for Murray Gell-Mann, whom Berkeley was trying to hire. It
was felt that students would be easier to dislodge than a faculty
member. Gross and I wrote one joint paper in 1965 \cite{Gross:1965},
which I felt was rather clever.}

Geoffrey Chew's approach to understanding the strong interactions
was based on several general principles
\cite{Chew:1961a,Chew:1966a}. He was very persuasive in advocating
them, and I was strongly influenced by him. The first principle was
that quantum field theory, which was so successful in describing
QED, was inappropriate for describing a strongly interacting theory,
where a weak-coupling perturbation expansion would not be useful. A
compelling reason for holding this view was that none of the hadrons
(particles that have strong interactions) seemed to be more
fundamental than any of the others. Therefore a field theory that
singled out some subset of the hadrons did not seem sensible. Also,
it was clearly not possible to formulate a quantum field theory with
a fundamental field for every hadron. One spoke of {\em nuclear
democracy} to describe this situation.\footnote{The quark concept
arose during this period, but the prevailing opinion was that quarks
are just mathematical constructs. The SLAC deep inelastic scattering
experiments in the late 1960s made it clear that quarks and gluons
are physical (confined) particles. It was then natural to try to
base a quantum field theory on them, and QCD was developed a few
years later with the discovery of asymptotic freedom.}

For these reasons, Chew argued that field theory was inappropriate
for describing strong nuclear forces. Instead, he advocated
focussing attention on physical quantities, especially the S Matrix,
which describes on-mass-shell scattering amplitudes.  The goal was
therefore to develop a theory that would determine the S matrix.
Some of the ingredients that went into this were properties deduced
from quantum field theory, such as unitarity and maximal analyticity
of the S matrix. These basically encode the requirements of
causality and nonnegative probabilities.

Another important proposal, due to Chew and Frautschi, whose
necessity was less obvious, was maximal analyticity in angular
momentum \cite{Chew:1961yz,Chew:1962eu}. The idea is that partial
wave amplitudes $a_l (s)$, which are defined in the first instance
for angular momenta $l=0,1,\ldots$, can be uniquely extended to an
analytic function of $l$, $a(l,s)$, with isolated poles called Regge
poles. The Mandelstam invariant $s$ is the square of the invariant
energy of the scattering reaction. The position of a Regge pole is
given by a Regge trajectory $l = \a(s)$. The values of $s$ for which
$l$ takes a physical value, correspond to physical hadron states.
The necessity of branch points in the $l$ plane, with associated
Regge cuts, was established by Mandelstam. Their role in
phenomenology was less clear.

The theoretical work in this period was strongly influenced by
experimental results. Many new hadrons were discovered in
experiments at the Bevatron in Berkeley, the AGS in Brookhaven, and
the PS at CERN. Plotting masses squared versus angular momentum (for
fixed values of other quantum numbers), it was noticed that the
Regge trajectories are approximately linear with a common slope
$$
\a(s) = \a(0) + \a' s \qquad \qquad \a' \sim 1.0\, ({\rm
GeV})^{-2}\, .
$$
Using the crossing-symmetry properties of analytically continued
scattering amplitudes, one argued that exchange of Regge poles (in
the $t$ channel) controlled the high-energy, fixed momentum
transfer, asymptotic behavior of physical amplitudes:
$$
A(s, t) \sim \b(t) (s/s_0)^{\a(t)} \qquad s \to \infty, \, t <0.
$$
In this way one deduced from data that the intercept of the $\r$
trajectory, for example, was $\a_\r(0) \sim .5$. This is consistent
with the measured mass $m_\r =.76 \, {\rm GeV}$ and the Regge slope
$\a' \sim 1.0\, ({\rm GeV})^{-2}$.

The ingredients discussed above are not sufficient to determine the
S matrix, so one needed more. Therefore, Chew advocated another
principle called the {\em bootstrap}. The idea was that the exchange
of hadrons in crossed channels provide forces that are responsible
for causing hadrons to form bound states. Thus, one has a
self-consistent structure in which the entire collection of hadrons
provides the forces that makes their own existence possible. It was
unclear for some time how to formulate this intriguing property in a
mathematically precise way. As an outgrowth of studies of {\em
finite-energy sum rules} in 1967
\cite{Dolen:1967,Igi:1967a,Igi:1967b,Logunov:1967,Dolen:1968} this
was achieved in a certain limit in 1968
\cite{Freund:1968,Harari:1969,Rosner:1969}. The limit, called the
{\em narrow resonance approximation} was one in which resonance
lifetimes are negligible compared to their masses. The observed
linearity of Regge trajectories suggested this approximation, since
otherwise pole positions would have significant imaginary parts. In
this approximation branch cuts in scattering amplitudes, whose
branch points correspond to multiparticle thresholds, are
approximated by a sequence of resonance poles.

The bootstrap idea had a precise formulation in the narrow resonance
approximation, which was called {\em duality}. This is the statement
that a scattering amplitude can be expanded in an infinite series of
$s$-channel poles, and this gives the same result as its expansion
in an infinite series of $t$-channel poles.\footnote{One defines
divergent series by analytic continuation.} To include both sets of
poles, as usual Feynman diagram techniques might suggest, would
amount to double counting.

\section{1968 -- 70: The dual resonance model}

I began my first postdoctoral position (entitled {\em instructor})
at Princeton University in 1966. For my first two and a half years
there, I continued to do work along the lines described in the
previous section (Regge pole theory, duality, etc.). Then Veneziano
dropped a bombshell -- an exact analytic formula that exhibited
duality with linear Regge trajectories \cite{Veneziano:1968}.
Veneziano's formula was designed to give a good phenomenological
description of the reaction $\pi + \pi \to \pi + \omega$ or the
decay $\omega \to \pi^+ + \pi^0 + \pi^-$. Its structure was the sum
of three Euler beta functions:
$$
T = A(s,t) + A(s, u) + A(t,u)
$$
$$
A(s,t) = \frac{\G(-\a(s)) \G(- \a(t))}{\G(-\a(s) -\a(t)) },
$$
where $\a$ is a linear Regge trajectory
$$
\a(s) = \a(0) + \a' s .
$$
An analogous formula appropriate to the reaction $\pi + \pi \to \pi
+ \pi$ was quickly proposed by Lovelace and Shapiro
\cite{Lovelace:1968,Shapiro:1969}. A rule for building in adjoint
$SU(N)$ quantum numbers was formulated by Chan and Paton
\cite{Paton:1969}. This symmetry was initially envisaged to be a
global (flavor) symmetry, but it later turned out to be a local
gauge symmetry.

The Veneziano formula gives an explicit realization of duality and
Regge behavior in the {\em narrow resonance approximation}. The
function $A(s,t)$ can be expanded in terms of the $s$-channel poles
or the $t$-channel poles. The motivation for writing down this
formula was mostly phenomenological, but it turned out that formulas
of this type describe tree amplitudes in a perturbatively consistent
quantum theory!

Very soon after the appearance of the Veneziano amplitude, Virasoro
proposed an alternative formula \cite{Virasoro:1969a}
$$
T = \frac{\G(-\half \a(s)) \G(- \half\a(t))
\G(- \half\a(u))}{\G(-\half\a(t) -\half\a(u)) \G(-\half\a(s)
-\half\a(u))\G(-\half\a(s) -\half\a(t)) },
$$
which has similar virtues. Since this formula has total $stu$
symmetry, it is only applicable to particles that are singlets of
the Chan--Paton group.

Over the course of the next year or so, string theory (or {\em dual
models}, as the subject was then called) underwent a sudden surge of
popularity, marked by several remarkable discoveries. One was the
discovery of an $N$-particle generalization of the Veneziano formula
\cite{Bardakci:1969a,Goebel:1969,Chan:1969b,Koba:1969a,Koba:1969b}:
$$
A_N (k) = g_{\rm open}^{N-2} \int d\mu_N(y)  \prod_{i<j} (y_i -
y_j)^{\a' k_i \cdot k_j},
$$
where $y_1, y_2, \ldots, y_N$ are real coordinates, any three of
which are $y_A, y_B, y_C$, and
$$
d\mu_N(y) = |(y_A  -y_B)(y_B -y_C)(y_C-y_A)| \prod_{i=1}^{N-1}
\theta(y_{i+1} - y_i)
$$
$$
\times \d(y_A-y_A^0) \d(y_B-y_B^0) \d(y_C-y_C^0) \prod_{i=1}^N d
y_i.
$$
The formula is independent of $y_A^0,y_B^0,y_C^0$ due to its
$SL(2,\IR)$ symmetry, which allows them to be mapped to arbitrary
real values. This formula has cyclic symmetry in the $N$ external
lines.

Soon thereafter Shapiro formulated an $N$-particle generalization of
the Virasoro formula \cite{Shapiro:1970}:
$$
A_N(k_1,k_2,\dots,k_N) = g_{\rm closed}^{N-2} \int
d\mu_N(z)\prod_{i<j} |z_i-z_j|^{\a'k_i\cdot k_j } ,
$$
where $z_1, z_2, \ldots, z_N$  are complex coordinates, any three of
which are $z_A, z_B, z_C$,  and
$$
d\mu_N(z) = |(z_A -z_B)(z_B -z_C)(z_C-z_A)|^2
$$
$$
\times \d^2(z_A-z_A^0) \d^2(z_B-z_B^0) \d^2(z_C-z_C^0) \prod_{i=1}^N
d^2 z_i .
$$
The formula is independent of $z_A^0,z_B^0,z_C^0$ due to its
$SL(2,\IC)$ symmetry, which allows them to be mapped to arbitrary
complex values. This amplitude has total symmetry in the $N$
external lines.

Both of these formulas were shown to have a consistent factorization
on a spectrum of single-particle states described by an infinite
number of harmonic oscillators
\cite{Fubini:1969a,Fubini:1969b,Bardakci:1969b,Fubini:1970,Nambu:1970a}
$$ \{ a^\m_m \} \qquad \m = 0,1,\dots,d-1 \qquad m= 1, 2, \ldots $$
with one set of such oscillators in the Veneziano case and two sets
in the Virasoro case. These results were interpreted as describing
the scattering of modes of a relativistic string
\cite{Nambu:1970a,Nambu:1970b,Susskind:1970e,Susskind:1970f,Nielsen:1970a,Fairlie:1970}:
open strings in the first case and closed strings in the second
case. Amazingly, the formulas preceded the interpretation. Although,
we did not propose a string interpretation, Gross, Neveu, Scherk,
and I did realize that the relevant diagrams of the loop expansion
were classified by the possible topologies of two-dimensional
manifolds with boundaries \cite{Gross:1970a}.

Having found the factorization, it became possible to compute
radiative corrections (loop amplitudes). This was initiated by
Kikkawa, Sakita, and Virasoro \cite{Kikkawa:1969} and followed up by
many others. Let me describe my role in this. I was at Princeton,
where I collaborated with Gross, Neveu, and Scherk in computing
one-loop amplitudes. In particular, we discovered unanticipated
singularities in the ``nonplanar'' open-string loop diagram
\cite{Gross:1970b}. The world sheet is a cylinder with two external
particles attached to each boundary. Our computations showed that
this diagram gives branch points that violate unitarity. This was a
very disturbing conclusion, since it seemed to imply that the
classical theory does not have a consistent quantum extension. This
was also discovered by Frye and Susskind \cite{Frye:1970b}. (The
issue of quantum consistency turned out to be a recurring theme,
which reappeared many years later, as discussed in Section 7.)

Soon thereafter Claude Lovelace pointed out \cite{Lovelace:1971}
that these branch points become poles provided that
$$
\a(0)=1 \quad \and \quad d=26.
$$
Until Lovelace's work, everyone assumed that the spacetime dimension
was $d=4$.\footnote{The idea of considering a higher dimension was
suggested to Lovelace by David Olive.} As we were not yet talking
about gravity, there was no reason to consider anything else. Later,
these poles were interpreted as closed-string modes in a one-loop
open-string amplitude. Nowadays this is referred to as open
string--closed string duality.

Lovelace's analysis also required there to be an infinite number of
decoupling conditions. These turned out to be precisely the Virasoro
constraints, which were discovered at about the same time
\cite{Virasoro:1970,Fubini:1971}. A couple of years later Brink and
Olive constructed a physical-state projection operator
\cite{Brink:1973qm}, which they used to verify Lovelace's conjecture
that the nonplanar loop amplitude actually contains closed-string
poles when the decoupling conditions in the critical dimension are
imposed \cite{Brink:1973gi}.

Thus, quantum consistency was restored, but the price was high: a
spectrum with a tachyon and 22 extra dimensions of space. In 1973,
the origin of the critical dimension and the intercept condition
were explained in terms of the light-cone gauge quantization of a
fundamental string by Goddard, Goldstone, Rebbi, and Thorn
\cite{Goddard:1973}. Prior to this paper the string interpretation
of dual models was only a curiosity. The GGRT approach was extended
to interacting strings by Mandelstam \cite{Mandelstam:1973}.

\section{1971 -- 73: The RNS model}

In January 1971 Pierre Ramond constructed a dual-resonance model
generalization of the Dirac equation \cite{Ramond:1971b}. He
reasoned as follows: just as the total momentum of a string, $p^\m$,
is the zero mode of a momentum density $P^\m (\s)$, so should the
Dirac matrices $\g^\m$ be the zero modes of densities $\G^\m (\s)$.
Then he defined the modes of $\G \cdot P$:
$$ F_n = \int_0^{2\pi} e^{-in\s} \G \cdot P d\s \qquad n\in \IZ.$$
In particular,
$$ F_0 = \g \cdot p + {\rm oscillator \, terms}\, .$$
He proposed the wave equation
$$ (F_0 + m) |\psi\rangle =0,$$
which is now known as the {\em Dirac--Ramond Equation}. Its
solutions give the spectrum of a noninteracting fermionic string.

Ramond also observed that the Virasoro algebra generalizes
to\footnote{His paper does not include the central terms.}
$$ \{ F_m, F_n \} = 2 L_{m+n} + \frac{c}{3} m^2 \d_{m,-n}$$
$$ [ L_m, F_n ] = (\frac{m}{2} -n) F_{m+n}$$
$$ [L_m, L_n] = (m-n) L_{m+n} + \frac{c}{12} m^3 \d_{m,-n}\, .$$
The free fermion spectrum should be restricted by the super-Virasoro
constraints $F_n |\psi\rangle = L_n |\psi\rangle =0$ for $n>0$.

Andr\'e Neveu and I proposed a new bosonic dual model, which we
called the {\em dual pion model,} in March 1971
\cite{Neveu:1971b}.\footnote{We submitted another publication
\cite{Neveu:1971fz} one month earlier that contained some, but not
all, of the right ingredients.} It has a similar structure to
Ramond's free fermion theory, with the periodic density $\G^\m(\s)$
replaced by an antiperiodic one $H^\m (\s)$. Then the modes
$$ G_r = \int_0^{2\pi} e^{-ir\s} H \cdot P d\s \qquad r \in \IZ + 1/2$$
satisfy a similar super-Virasoro algebra. The free particle spectrum
is given by the wave equation $(L_0 -1/2)|\psi\rangle =0$
supplemented by the constraints $ G_r |\psi\rangle =0$ for $r>0$.
(These formulas are appropriate in the ${\cal F}_2$ picture
discussed below.) We also constructed $N$-particle amplitudes
analogous to those of the Veneziano model.

The $\pi + \pi \to \pi + \pi$ amplitude computed in the dual pion
model turned out to have exactly the form that had been proposed
earlier by Lovelace and Shapiro. However, the intercepts of the
$\pi$ and $\rho$ Regge trajectories were $\a_{\pi}(0) = 1/2$ and
$\a_{\rho}(0) = 1$. These were half a unit higher than was desired
in each case. This implied that the pion was tachyonic and the rho
was massless.

Soon after our paper appeared, Neveu traveled to Berkeley, where
there was considerable interest in our results. This led to Charles
Thorn (a student of Stanley Mandelstam at the time) joining us in a
follow-up project in which we proved that the super-Virasoro
constraints were fully implemented \cite{Neveu:1971c}. This required
recasting the original description of the string spectrum (called
the ${\cal F}_1$ picture) in a new form, which we called the ${\cal
F}_2$ picture. The three of us then assembled these bosons together
with Ramond's fermions into a unified interacting theory of bosons
and fermions \cite{Neveu:1971d,Thorn:1971}, thereby obtaining an
early version of what later came to be known as superstring theory.

The string world-sheet theory that gives this spectrum of bosons and
fermions is
$$
S = \int d\s d\t \left( \pa_\a X^\m \pa^\a X_\m -i \bar\psi^\m \r^\a
\pa_\a \psi_\m\right),
$$
where $\psi^\m$ are two-dimensional Majorana spinors and $\r^\a$ are
two-dimensional Dirac matrices. Later in 1971 Gervais and Sakita
observed \cite{Gervais:1971b} that this action has {\em
two-dimensional global supersymmetry} described by the infinitesimal
fermionic transformations
$$
\d X^\m = \bar\e \psi^\m
$$
$$
\d\psi^\m = -i \r^\a \e \pa_\a X^\m .
$$
This is actually part of a much larger {\em superconformal
symmetry}. There are two possible choices of boundary conditions for
the fermi fields $\psi^\m$, one of which gives the boson spectrum
(Neveu--Schwarz sector) and the other of which gives the fermion
spectrum (Ramond sector).
Five years later, a more fundamental world-sheet action with local
supersymmetry was discovered \cite{Brink:1976,Deser:1976}. It has
the virtue of accounting for the super-Virasoro constraints as
arising from covariant gauge fixing.

Bruno Zumino explored the RNS string's gauge conditions associated
to the two-dimensional superconformal algebra \cite{Zumino:1974}.
Following that, he and Julius Wess began to consider the possibility
of constructing four-dimensional field theories with analogous
features. This resulted in their famous work \cite{Wess:1974} on
globally supersymmetric field theories in four dimensions. As a
consequence of their paper,\footnote{The work of Golfand and
Likhtman \cite{Golfand:1971iw}, which was the first to introduce the
four-dimensional super-Poincar\'e group, was not known in the West
at that time.} supersymmetry quickly became an active research
topic.

The dual pion model has a manifest $\IZ_2$ symmetry. Since the pion
is odd and the rho is even, this symmetry was identified with G
parity.\footnote{G parity is a hadronic symmetry that is a
consequence of charge conjugation invariance and isotopic spin
symmetry.} It was obvious that one could make a consistent
truncation (at least at tree level) to the even G-parity sector and
that then the model would be tachyon free. Because of the desired
identification with physical hadrons, there was no motivation (at
the time) to do that. Rather, considerable effort was expended in
the following years attempting to modify the model so as to lower
the intercepts by half a unit. As discussed in the next section,
none of these constructions was entirely satisfactory.

One of the important questions in this period was whether all the
physical string excitations have a positive norm. States of negative
norm (called {\em ghosts}) would represent a breakdown of unitarity
and causality, so it was essential that they not be present in the
string spectrum. The first proof of the {\em no-ghost theorem} for
the original bosonic string theory was achieved by Brower
\cite{Brower:1972}, building on earlier work by Del Giudice, Di
Vecchia, and Fubini \cite{DelGiudice:1972}. This work showed that a
necessary condition for the absence of ghosts is $d \leq 26$, and
that the critical value $d=26$ has especially attractive features,
as we already suspected based on the earlier observations of
Lovelace.

I generalized Brower's proof of the no-ghost theorem to the RNS
string theory and showed that $d=10$ is the critical dimension and
that the ground state fermion should be massless
\cite{Schwarz:1972}. This was also done by Brower and Friedman a bit
later \cite{Brower:1973}. An alternative, somewhat simpler, proof of
the no-ghost theorem for both of the string theories was given by
Goddard and Thorn at about the same time \cite{Goddard:1972}. Other
related work included \cite{Gervais:1973,Olive:1973,Corrigan:1974}.

Later in 1972, thanks to the fact that Murray Gell-Mann had become
intrigued by my work with Neveu, I was offered a senior research
appointment at Caltech. I think that the reason Gell-Mann became
aware of our work was because he spent the academic year 1971-72 on
a sabbatical at CERN, where there was an active dual models group. I
felt very fortunate to receive such an offer, especially in view of
the fact that the job market for theoretical physicists was
extremely bad at the time. Throughout the subsequent years at
Caltech, when my work was far from the mainstream, and therefore not
widely appreciated, Gell-Mann was always very supportive. For
example, he put funds at my disposal to invite visitors. This
facilitated various collaborations with Lars Brink, Jo\"el Scherk,
and Michael Green among others.

One of the first things I did at Caltech was to study the
fermion-fermion scattering amplitude. Using the physical-state
projection operator \cite{Brink:1973qm}, Olive and Scherk had
derived a formula that involved the determinant of an infinite
matrix \cite{Olive:1974sv}. C.C. Wu and I \cite{Wu:1973} discovered
that this determinant is a simple function. We derived the result
analytically in a certain limit and then verified numerically that
it is exact everywhere. (The result was subsequently verified
analytically \cite{Corrigan:1974vz}.) To our surprise, the
fermion-fermion scattering amplitude ended up looking very similar
to the bosonic amplitudes. This might have been interpreted as a
hint of spacetime supersymmetry, but this was before the
Wess--Zumino paper, and that was not yet on my mind.

String theory is formulated as an on-shell S-matrix theory in
keeping with its origins discussed earlier. However, the SLAC deep
inelastic scattering experiments in the late 1960s made it clear
that the hadronic component of the electromagnetic current is a
physical off-shell quantity, and that its asymptotic properties
imply that hadrons have hard pointlike constituents. With this
motivation, I tried for the next year or so to construct off-shell
amplitudes. Although some intriguing results were obtained
\cite{Schwarz:1974a,Schwarz:1974b,Schwarz:1974c}, this was
ultimately unsuccessful. Moreover, all indications were that strings
were too soft to describe hadrons with their pointlike constituents.

At this point there were many good reasons to stop working on string
theory: a successful and convincing theory of hadrons (QCD) was
discovered, and string theory had many severe problems as a hadron
theory. These included an unrealistic spacetime dimension, an
unrealistic spectrum, and the absence of pointlike constituents.
Also, convincing theoretical and experimental evidence for the
standard model was rapidly falling into place. Understandably, given
these successes and string theory's shortcomings, string theory
rapidly fell out of favor. What had been a booming enterprise
involving several hundred theorists rapidly came to a grinding halt.

Given that the world-sheet descriptions of the two known string
theories have conformal invariance and superconformal invariance, it
was a natural question whether one could obtain new string theories
described by world-sheet theories with extended superconformal
symmetry. The $N=2$ case was worked out in \cite{Ademollo:1976pp}.
The critical dimension is four, but the signature has to be $(2,2)$.
For a long time it was believed that the critical dimension of the $N=4$
string is negative, but in 1992 Siegel argued that (due to the reducibility
of the constraints) the $N=4$ string is the same as the $N=2$ string
\cite{Siegel:1992ev}.

\section{1974 -- 75: Gravity and unification}

The string theories that were known in the 1970s (the bosonic string
and the RNS string) had many shortcomings as a theory of
hadrons. The most obvious of these was the necessity of an
unrealistic spacetime dimension (26 or 10). Another is the
occurrence of tachyons in the spectrum, which implies that the
vacuum is unstable. However, the one that bothered us the most
was the presence of massless particles in the spectrum, which
do not occur in the hadron spectrum.

In both string theories the spectrum of open strings contains
massless spin 1 particles, and the spectrum of closed strings
contains a massless spin 2 particle as well as other massless
particles (a dilaton and an antisymmetric tensor in the case of
oriented bosonic strings). These particles lie on the leading Regge
trajectories in their respective sectors. Thus, the leading
open-string Regge trajectory has intercept $\a(0)=1$, and the
leading closed-string Regge trajectory has intercept $\a(0)=2$. Is
was tempting (in the RNS model) to identify the leading open-string
trajectory as the one for the $\rho$ meson. In fact, it had most of
the properties expected for that case except that the empirical
intercept is about $\a_{\rho}(0)=1/2$. The leading closed-string
trajectory carries vacuum quantum numbers, as expected for the {\em
Pomeron} trajectory. Moreover, the factor of two between the
open-string and closed-string Regge slopes is approximately what is
required by the data. However, the Pomeron intercept was also double
of the desired value $\a(0)=1$, which is the choice that could
account for the near constancy (up to logarithmic corrections) of
hadronic total cross sections at high energy.

For these reasons we put considerable effort in the years 1972--74
into modifying the RNS theory in such a way as to lower all
open-string Regge trajectories by half a unit and all closed-string
Regge trajectories by one unit. Some successes along these lines
actually were achieved, accounting for some aspects of chiral
symmetry and current algebra \cite{Neveu:1971ma,Schwarz:1972gr}.
However, none of the schemes was entirely consistent. The main
problem is that changing the intercepts and the spacetime dimensions
meant that the Virasoro constraints were not satisfied, and so the
spectrum was not ghost-free. Also, the successes for nonstrange
mesons did not extend to mesons made from heavy quarks, and the
Ramond fermions didn't really look like baryons.

The alternative to modifying string theory to get what we wanted was
to understand better what the theory was giving without modification.
String theories in the critical dimension clearly were beautiful
theories, with a remarkably subtle and intricate structure,
and they ought to be good for something. The fact that
they were developed in an attempt to understand hadron physics did
not guarantee that this was necessarily their appropriate physical
application. Furthermore, the success of QCD made the effort to
formulate a string theory of hadrons less pressing.

The first indication that such an agnostic attitude could prove
worthwhile was a pioneering work by Neveu and Scherk
\cite{Neveu:1972}, which studied the interactions of the massless
spin 1 open-string particles at low energies (or, equivalently, in
the {\em zero-slope limit}) and proved that their interactions
agreed with those of Yang--Mills gauge particles in the adjoint
representation of the Chan--Paton group.\footnote{This work
relating string theory and Yang--Mills theory followed an earlier
study by Scherk describing how to obtain $\phi^3$ field theory in
the zero-slope limit \cite{Scherk:1971xy}.} In other words, open-string
theory was Yang--Mills gauge theory modified by higher dimension
interactions at the string scale. This implies that the Chan--Paton
group is actually a Yang--Mills gauge group.  Prior to this work by
Neveu and Scherk, it was always assumed that the Chan--Paton symmetry
is a global symmetry. I am struck
by the fact that Yang and Mills in their original paper on SU(2)
gauge theory \cite{Yang:1954ek}, tried to identify
the gauge symmetry with isotopic spin symmetry
and the gauge fields with $\rho$ mesons. In our failed efforts to
describe hadrons, we had been making essentially the same mistake.

I arranged for Jo\"el Scherk, with whom I had collaborated in
Princeton, to visit Caltech in the winter and spring of 1974. Our
interests and attitudes in physics were very similar, and so we were
anxious to start a new collaboration. Each of us felt that string
theory was too beautiful to be just a mathematical curiosity. It
ought to have some physical relevance. We had frequently been struck
by the fact that string theories exhibit unanticipated miraculous
properties. What this means is that they have a very deep
mathematical structure that is not fully understood. By digging
deeper one could reasonably expect to find more surprises and then
learn new lessons. Therefore, despite the fact that the rest of the
theoretical high energy physics community was drawn to the important
project of exploring the standard model, we wanted to explore string
theory.

Since my training was as an elementary particle physicist, gravity
was far from my mind in early 1974. Traditionally, elementary
particle physicists had ignored the gravitational force, which is
entirely negligible under ordinary circumstances. For these reasons,
we were not predisposed to interpret string theory as a physical
theory of gravity. General relativists, the people who did study
gravity, formed a completely different community. They attended
different meetings, read different journals, and had no need for
serious communication with particle physicists, just as particle
physicists felt they had no need for relativists who studied topics
such as black holes or the early universe.

Despite all this, we decided to do what could have been done two
years earlier: we explored whether it is possible to interpret the
massless spin 2 state in the closed-string spectrum as a graviton.
This required carrying out an analysis analogous to the earlier one
of Neveu and Scherk. This time one needed to decide whether the
interactions of the massless spin 2 particle in string theory agree
at low energy with those of the graviton in general relativity (GR).
Success was inevitable, because GR is the only consistent
possibility at low energies (\ie neglecting corrections due to
higher-dimension operators), and critical string theory certainly is
consistent. At least, it contains the requisite gauge invariances
to decouple all but the transverse polarizations.
Therefore, the harder part of this work was forcing
oneself to ask the right question. Finding the right answer was easy.
In fact, by invoking certain general theorems, due to
Weinberg \cite{Weinberg:1965rz}, we were able
to argue that string theory agrees with general relativity at low
energies \cite{Scherk:1974a}. Although we were not aware of it at
the time, Tamiaki Yoneya had obtained the same result somewhat
earlier \cite{Yoneya:1973,Yoneya:1974}.

In \cite{Scherk:1974a}, Scherk and I proposed to interpret string theory as a
quantum theory of gravity, unified with the other forces. This meant
taking the whole theory seriously, not just viewing it as a framework for
deriving GR and Yang--Mills theory as limits. Our paper was entitled
{\em Dual Models for Non-Hadrons}, which I think was a poor choice.
It emphasized the fact that we were no longer trying to describe
hadrons and their interactions, but it failed to emphasize what we
were proposing to do instead. A better choice would have been
{\em String Theory as a Quantum Theory of Gravity Unified with the
Other Forces}.

This proposal had several advantages: First, gravity was required by
both of the known critical string theories. a forceful way of
expressing this is {\em the assumption that the fundamental physical
entities are strings predicts the existence of gravity}. In fact,
even today, this is really the only direct experimental evidence
that exists in support of string theory, though there are many other
reasons to take string theory seriously.

Second, string theories are free from the UV divergences that
typically appear in point-particle theories of gravity. The reason
can be traced to the extended structure of strings. Specifically,
string world sheets are smooth, even when they describe
interactions. So they do not have the short-distance singularities
that are responsible for UV divergences. These divergences always
occur when one attempts to interpret general relativity (with or
without matter) as a quantum field theory, since ordinary Feynman
diagrams do have short-distance singularities.\footnote{There has
been recent speculation about the possible finiteness of ${\cal N} =
8$ supergravity. If true, this would be a counterexample to my
assertion. Whether or not this is the case, the  nonperturbative
completion of ${\cal N} = 8$ supergravity is likely to lead one back
to ten-dimensional superstring theory \cite{Green:2007zzb}.} This
can also be expressed forcefully by saying {\em the assumption that
the fundamental physical entities are point particles predicts that
gravity does not exist!} I found this line of reasoning very
compelling.\footnote{In light of subsequent developments,
the distinction made here no longer seems quite so sharp.
A single theory can have dual descriptions
based on different fundamental entities. One description is weakly
coupled when the other one is strongly coupled. For example, certain
AdS/CFT duality relates string theory in a particular background
geometry to a more conventional quantum field theory.}

Third, extra dimensions could be a very good thing, rather than a
problem, since in a gravity theory the geometry of spacetime is
determined by the dynamics. (Prior to our proposal,
their appearance as critical dimensions of string theories
was viewed as a shortcoming of the theories rather than as
a reason to study their possible implications.)
In the gravitational setting one could imagine that the equations of
motion would require (or at least allow) the extra dimensions to
form a very small compact manifold. In 1974 there had been
essentially no work on Kaluza--Klein theory for many years (and
certainly none in the particle physics community) at the time of
this work, so the notion of extra dimensions seemed very bizarre
to most particle theorists.
It is discussed so much nowadays that it is easy to forget
this fact.  Since the only length scale in string theory is the string scale,
determined by the string tension,
that would be the natural first guess for the size of the compact
space. Given this assumption, the value of Newton's constant in four
dimensions can be deduced. The observed strength of gravity requires
a Regge slope $\a' \sim 10^{-38} \, {\rm GeV}^{-2}$ instead of $\a'
\sim 1 \, {\rm GeV}^{-2}$, which is the hadronic value. Thus, the
change in interpretation meant that the tension of the strings,
which is proportional to the reciprocal of $\a'$, needed to be
increased by 38 orders of magnitude. Equivalently, the size of the
strings decreased by 19 orders of magnitude. This was a big conceptual
leap, though the mathematics was unchanged.

Fourth, unification of gravity with other forces described by
Yang--Mills theories was automatic when open strings are included.
Of course, it was immediately clear that the construction of a
realistic vacuum would be a great challenge. Indeed, that is
where much of the effort these days is focussed. Other ways
of incorporating gauge interactions in string theory were
discovered many years later. These include heterotic string theory,
where gauge fields appear as closed-string modes in ten dimensions,
coincident D-brane world-volume theories, and certain types
of singularities in M theory or F theory.

Scherk and I were very excited by the possibility that string theory
could be the Holy Grail of unified field theory, overcoming the
problems that had stymied other approaches. In addition to
publishing our work in scholarly journals, we gave numerous lectures
at conferences and physics departments all over the world. We even
submitted a paper entitled {\em Dual Model Approach to a Renormalizable
Theory of Gravitation} to the 1975 essay competition of the Gravity
Research Foundation \cite{Scherk:1975a}.\footnote{It would
have been better to say {\em ultraviolet finite} instead of
{\em renormalizable}.} The first paragraph of that
paper reads as follows:

{\it A serious shortcoming of Einstein's theory of gravitation is the
nonrenormalizability of its quantum version when considered in interaction
with other quantum fields. In our opinion this is a genuine problem
requiring a modification of the theory. It is suggested in this essay that
dual resonance models may provide a suitable framework for such a
modification, while at the same time achieving a unification with
other basic interactions.}

For the most part our work
was received politely --- as far as I know, no one accused us of
being crackpots. Yet, for a decade, very few experts showed
much interest. Part of the problem may have been that some key people
were unaware of our proposal. Unfortunately, Scherk passed away midway
through this 10-year period, though not before making some other
important contributions that are discussed in the following
sections. In the decade following its publication, our paper
\cite{Scherk:1974a} only received about 20 citations in papers
written by people other than Scherk or myself. The authors of these
papers include the following distinguished physicists: Lars Brink,
Peter Freund, Michael Green, Bernard Julia,
David Olive, Tamiaki Yoneya, and Bruno Zumino. After the subject
took off in the autumn of 1984, \cite{Scherk:1974a} became much
better known.

\section{1975 -- 79: Supersymmetry and supergravity}

Following the pioneering work of Wess and Zumino, discussed earlier,
the study of supersymmetric quantum field theories became a major
endeavor. One major step forward was the realization that
supersymmetry can be realized as a local symmetry. This requires
including a gauge field, called the {\em gravitino field}, which is
vector-spinor. In four dimensions it describes a massless particle
with spin 3/2, which is the supersymmetry partner of the graviton.
Thus, local supersymmetry only appears in gravitational theories,
which are called supergravity theories.

The first example of a supergravity theory was ${\cal N} = 1$, $d=4$
supergravity. It was formulated in a second-order formalism by
Ferrara, Freedman, and Van Nieuwenhuizen \cite{Freedman:1976} and
subsequently in a first-order formalism by Deser and Zumino
\cite{Deser:1976eh}. The first-order formalism simplifies the
analysis of terms that are quartic in fermi fields.

The two-dimensional locally supersymmetric and reparametrization
invariant formulation of the RNS world-sheet action was constructed
very soon thereafter \cite{Brink:1976,Deser:1976}.\footnote{This
generalized the one-dimensional result obtained a bit earlier for a
spinning point particle \cite{Brink:1976sz}.} This construction was
generalized to the $N=2$ string of \cite{Ademollo:1976pp} by Brink
and me \cite{Brink:1976vg}. Reparametrization-invariant world-sheet
actions of this type are frequently associated with the name
Polyakov, because he used them very skillfully five years later in
constructing the path-integral formulation of string theory
\cite{Polyakov:1981rd,Polyakov:1981re}. Since neither Polyakov nor
the authors of \cite{Brink:1976,Deser:1976} are happy with this
usage, the new textbook \cite{Becker:2007cup} refers to this type of
world-sheet action as a {\em string sigma-model action}.

The RNS closed-string spectrum contains a massless gravitino (in ten
dimensions) in addition to the graviton discussed in the previous
section.\footnote{More precisely, as was understood later, there are
one or two gravitinos depending on whether one is describing a type
I or type II superstring.} Since this is a gauge field, the only way
the theory could be consistent is if the theory has local
supersymmetry. This requires, in particular, that the spectrum
should contain an equal number of bosonic and fermionic degrees of
freedom at each mass level. However, as it stood, this was not the
case. In particular, the bosonic sector contained a tachyon (the
``pion''), which had no fermionic partner.

In 1976 Gliozzi, Scherk, Olive \cite{Gliozzi:1976,Gliozzi:1977}
proposed a projection of the RNS spectrum -- {\em the GSO
Projection} -- that removes roughly half of the states (including
the tachyon). Specifically, in the bosonic (NS) sector they
projected away the odd G-parity states, a possibility that was
discussed earlier, and in the fermionic (R) sector they projected
away half the states, keeping only certain definite chiralities.
Then they counted the remaining physical degrees of freedom at each
mass level. After the GSO projection the masses of open-string
states, for both bosons and fermions, are given by $\a' M^2 =n$,
where $n=0,1,\ldots$ Denoting the open-string degeneracies of states
in the GSO-projected theory by $d_{\rm NS}(n)$ and $d_{\rm R}(n)$,
they showed that these are encoded in the generating functions
$$
f_{\rm NS}(w) = \sum_{n=0}^\infty d_{\rm NS}(n) w^n$$ $$=
\frac{1}{2\sqrt{w}} \left[\prod_{m=1}^{\infty}
\left(\frac{1+w^{m-1/2}}{1-w^m}\right)^8 - \prod_{m=1}^{\infty}
\left(\frac{1 -w^{m-1/2}}{1-w^m}\right)^8 \right].
$$
and
$$
f_{\rm R}(w) = \sum_{n=0}^\infty d_{\rm R}(n) w^n = 8
\prod_{m=1}^{\infty} \left(\frac{1+w^{m}}{1-w^m}\right)^8 .
$$

In 1829, Jacobi proved the remarkable identity \cite{Jacobi:1829}
$$
f_{\rm NS}(w) = f_{\rm R}(w),
$$
though he used a different notation. Thus, there are an equal
number of bosons and fermions at every mass level, as required. This
was compelling evidence (though not a proof) for {\em
ten-dimensional spacetime supersymmetry} of the GSO-projected
theory. Prior to this work, one knew that the RNS theory has
world-sheet supersymmetry, but the realization that the theory
should have spacetime supersymmetry was a major advance.

Since a Majorana--Weyl spinor in ten dimensions has 16 real
components, the minimal number of supercharges is 16. In particular,
the massless modes of open superstrings at low energies are
approximated by an ${\cal N} = 1$, $d=10$ super Yang--Mills theory
with 16 supersymmetries. This theory was constructed in
\cite{Gliozzi:1977,Brink:1977}. When this work was done, Brink and I
were at Caltech and Scherk was in Paris. Brink and I wrote to Scherk
informing him of our results and inviting him to join our
collaboration, which he gladly accepted. Brink and I were unaware of
the GSO collaboration, which was underway at that time, until their
work appeared. Both papers pointed out that maximally supersymmetric
Yang--Mills theories in less than ten dimensions could be deduced by
dimensional reduction, and both of them constructed the ${\cal N} =
4$, $d=4$ super Yang--Mills theory explicitly.

Having found the maximally supersymmetric Yang--Mills theories, it
was an obvious problem to construct the maximally supersymmetric
supergravity theories. Nahm showed \cite{Nahm:1977tg} that the
highest possible spacetime dimension for such a theory is $d=11$.
Soon thereafter, in a very impressive work, the Lagrangian for
${\cal N} = 1$, $d=11$ supergravity was constructed by Cremmer,
Julia, and Scherk \cite{Cremmer:1978}.
It was immediately clear that 11-dimensional supergravity is very
beautiful, and it aroused a lot of interest. However, it was
puzzling for a long time how it fits into the greater scheme of
things and whether it has any connection to string theory. Clearly,
supergravity in 11 dimensions is not a consistent quantum theory by
itself, since it is very singular in the ultraviolet. Moreover,
since superstring theory only has ten dimensions, it did not seem
possible that it could serve as a regulator. It took more than
fifteen years to find the answer to this conundrum
\cite{Townsend:1995kk,Witten:1995ex}: At strong coupling Type IIA
superstring theory develops a circular 11th dimension whose radius
grows with the string coupling constant. In the limit of infinite
coupling one obtains {\it M theory}, which is presumably a
well-defined quantum theory that has 11 noncompact dimensions.
Eleven-dimensional supergravity is the leading low-energy
approximation to M theory. In other words, M theory is the UV
completion of 11-dimensional supergravity.

In 1978--79, I spent the academic year at the \'Ecole Normale
Sup\'erieure in Paris supported by a Guggenheim Fellowship.
I was eager to work with Jo\"el Scherk on
supergravity, supersymmetrical strings, and related matters.
After various wide-ranging discussions we decided to focus on the
problem of supersymmetry breaking. We wondered how, starting from a
supersymmetric string theory in ten dimensions, one could end up
with a nonsupersymmetric world in four dimensions. The specific
supersymmetry breaking mechanism that we discovered can be explained
classically and does not really require strings, so we explored it
in a field theoretic setting \cite{Scherk:1979a,Scherk:1979b}.
The idea is that in a theory with
extra dimensions and global symmetries that do not commute with
supersymmetry ($R$ symmetries and $(-1)^F$ are examples), one could
arrange for a twisted compactification, and that this would break
supersymmetry. For example, if one extra dimension forms a circle,
the fields when continued around the circle could come could back
transformed by an R-symmetry group element. If the gravitino, in
particular, is transformed then it acquires mass in a consistent
manner.

An interesting example of our supersymmetry breaking mechanism was
worked out in a paper we wrote together with Eug\`ene Cremmer
\cite{Cremmer:1979}. We started with maximal supergravity in five
dimensions. This theory contains eight gravitinos that transform in
the fundamental representation of a USp(8) R-symmetry group. We took
one dimension to form a circle and examined the resulting
four-dimensional theory keeping the lowest Kaluza--Klein modes. The
supersymmetry-breaking R-symmetry element is a USp(8) element that
is characterized by four real mass parameters, since this group has
rank four. These four masses give the masses of the four complex
gravitinos of the resulting four-dimensional theory. In this way we
were able to find a consistent four-parameter deformation of ${\cal
N} = 8$ supergravity.

Even though the work that Jo\"el Scherk and I did on supersymmetry
breaking was motivated by string theory, we only discussed field theory
applications in our articles. The reason I never wrote about string
theory applications was that in the string theory setting it did not
seem possible to decouple the supersymmetry breaking mass parameters
from the compactification scales. This was viewed as a serious
problem, because the two scales are supposed to be hierarchically
different. In recent times, people have been considering string
theory brane-world scenarios in which much larger compactification
scales are considered. In such a context our supersymmetry breaking
mechanism might have a role to play. Indeed, quite a few authors
have explored various such possibilities.

\section{1979 -- 84: Superstrings and Anomalies}

Following Paris, I spent a month (July 1979) at CERN. There, Michael
Green and I unexpectedly crossed paths. We had become acquainted in
Princeton around 1970, when Green was at the IAS and I was at the
University, but we had not collaborated before. In any case,
following some discussions in the CERN cafeteria, we began a long
and exciting collaboration. Our first goal was to understand better
why the GSO-projected RNS string theory has spacetime supersymmetry.

Green, who worked at Queen Mary College London at the time, had
several extended visits to Caltech in the 1980--85 period, and I had
one to London in the fall of 1983. We also worked together several
summers in Aspen. On several of these occasions we also collaborated
with Lars Brink, who had visited Caltech and collaborated with me a
few times previously.

After a year or so of unsuccessful efforts, Green and I discovered a
new light-cone gauge formalism for the GSO-projected theory in which
spacetime supersymmetry of the spectrum and interactions was easily
proved. This was presented in three papers
\cite{Green:1980zg,Green:1981xx,Green:1981ya}. The first developed
the formalism, while the next two used this light-cone gauge
formalism to compute various tree and one-loop amplitudes and
elucidate their properties. At this stage only open-string
amplitudes were under consideration.

Our next project was to identify more precisely the possibilities
for superstring theories. The GSO work had identified the proper
projection for open strings, but it left unclear what one should do
with the closed strings. Green and I realized that there are three
distinct types of supersymmetry possible in ten dimensions and that
all three of them could be realized by superstring theories. In
\cite{Green:1981yb} we formulated the type I, type IIA, and type IIB
superstring theories. (We introduced these names a little later.)
The type I theory is a theory of unoriented open and closed strings,
whereas the type II theories are theories of oriented closed strings
only.

Brink, Green, and I formulated $d$-dimensional maximally
supersymmetric Yang--Mills theories and supergravity theories as
limits of superstring theory with $10-d$ of the ten dimensions
forming a torus. By computing one-loop string-theory amplitudes for
massless gauge particles in the type I theory and gravitons in the
type II theory and taking the appropriate limits, we showed that
both the Yang--Mills and supergravity theories are ultraviolet
finite at one loop for $d<8$ \cite{Green:1982sw}. The toroidally
compactified string-loop formulas exhibited T-duality symmetry,
though this was not pointed out explicitly in the article.

We also spent considerable effort formulating superstring field
theory in the light-cone gauge
\cite{Green:1982tc,Green:1983hw,Green:1984fu}. This work became
relevant about 20 years later, when the construction was generalized
to the case of type IIB superstrings in a plane-wave background
spacetime geometry.

The fact that our spacetime supersymmetric formalism was only
defined in the light-cone gauge was a source of frustration. Brink
and I had found a covariant world-line action for a massless
superparticle in ten dimensions
\cite{Brink:1981nb},\footnote{Casalbuoni
considered similar superparticle systems in four dimensions several
years earlier \cite{Casalbuoni:1976tz}.} so it was just a matter
of finding the suitable superstring generalization.  After a number
of attempts, Green and I eventually found a covariant world-sheet
action with manifest spacetime supersymmetry (and non-manifest kappa
symmetry) \cite{Green:1983wt,Green:1983sg}. This covariant action
reduces to our previous one in the light cone gauge, of course. It
was natural to try to use it to define covariant quantization.
However, due to a subtle combination of first-class and second-class
constraints, it was immediately apparent that this action is
extremely difficult to quantize covariantly. Numerous unsuccessful
attempts over the years bear testimony to the truth of this
assertion. More recently, Berkovits seems to have found a successful
scheme. However, as far as I can tell, its logical foundations are
not yet entirely clear.

Another problem of concern during this period was the formulation of
ten-dimensional type IIB supergravity,\footnote{Type IIA
supergravity can be obtained by dimensional reduction of
11-dimensional supergravity, but type IIB supergravity cannot be
obtained in this way.} which is the leading low-energy approximation
to type IIB superstring theory. Some partial results were obtained
in separate collaborations with Green \cite {Green:1982tk} and with
Peter West \cite{Schwarz:1983wa}. A challenging aspect of the
problem is the presence of a self-dual five-form field strength,
which obstructs a straightforward construction of a manifestly
covariant action. Therefore, I decided to focus on the equations of
motion, instead, which I presented in \cite{Schwarz:1983qr}.
Equivalent results were obtained in a superfield formalism by Howe
and West \cite{Howe:1983sr}.


Let me now turn to the issue of anomalies. Type I superstring theory
is a well-defined ten-dimensional theory at tree level for any $SO(n)$
or $Sp(n)$ gauge group \cite{Schwarz:1982md,Marcus:1982fr}. However, in
every case it is chiral (\ie parity violating) and the $d=10$ super
Yang--Mills sector is anomalous. Evaluation of a one-loop hexagon
diagram exhibits explicit nonconservation of gauge currents of the
schematic form
$$
\pa_\m J^\m \sim \e^{\m_1 \cdots \m_{10}} F_{\m_1 \m_2} \cdots
F_{\m_9 \m_{10}},
$$
which is a fatal inconsistency.

Alvarez-Gaum\'e and Witten derived general formulas for gauge,
gravitational, and mixed anomalies in an arbitrary spacetime
dimension \cite{AlvarezGaume:1983ig}, and they discovered that the
gravitational anomalies (nonconservation of the stress tensor)
cancel in type IIB supergravity. This result was not really a
surprise, since the one-loop type IIB superstring amplitudes are
ultraviolet finite. It appeared likely that type I superstring
theory is anomalous for any choice of the gauge group, but an
explicit computation was required to decide for sure. In this case
there are divergences that need to be regulated, so anomalies are
definitely possible.

Green and I explored the anomaly problem for type I superstring
theory off and on for almost two years until the crucial
breakthroughs were made in August 1984 at the Aspen Center for
Physics. That summer I was the organizer of a workshop entitled
``Physics in Higher Dimensions'' at the Aspen Center for Physics.
This attracted many participants, even though string theory was not
yet fashionable, because by that time there was considerable
interest in supergravity theories in higher dimensions and
Kaluza--Klein compactification. We benefitted from the presence of
many leading experts including Bruno Zumino, Bill Bardeen, Dan
Friedan, Steve Shenker, and others.

Green and I had tried unsuccessfully to compute the one-loop hexagon
diagram in type I superstring theory using our supersymmetric
light-cone gauge formalism, but this led to an impenetrable morass.
In discussions with Friedan and Shenker the idea arose to carry out
the computation using the covariant RNS formalism instead. At that
point, Friedan and Shenker left Aspen, so Green and I continued on
our own.

It soon became clear that both the cylinder and M\"obius-strip
diagrams contributed to the anomaly. Before a workshop seminar by
one of the other workshop participants (I don't remember which one),
I remarked to Green that there might be a gauge group for which the
two contributions cancel. At the end of the seminar Green said to me
``$SO(32)$,'' which was the correct result. Since this computation
only showed the cancellation of the pure gauge part of the anomaly,
we decided to explore the low-energy effective field theory to see
whether the gravitational and mixed anomalies could also cancel.
Before long, with the help of the results of Alvarez-Gaum\'e and
Witten and useful comments by Bardeen and others, we were able to
explain how this works. The effective field theory analysis was
written up first \cite{Green:1984sg}, and the string loop analysis
was written up somewhat later \cite{Green:1984qs}. We also showed
that the UV divergences of the cylinder and M\"obius-strip diagrams
cancel for $SO(32)$ \cite{Green:1984ed}. Nowadays such cancellations
are usually understood in terms of tadpole cancellations in a dual
closed-string channel.

The effective field theory analysis showed that $ E_8 \times E_8$ is
a second gauge group for which the anomalies could cancel for a
theory with ${\cal N} =1 $ supersymmetry in ten dimensions. In both
cases, it is crucial for the result that the coupling to
supergravity is included. The $SO(32)$ case could be accommodated by
type I superstring theory, but we didn't know a superstring theory
with gauge group $E_8 \times E_8.$ We were aware of the article by
Goddard and Olive that pointed out (among other things) that there
are just two even-self-dual Euclidean lattices in 16 dimensions, and
these are associated with precisely these two gauge groups
\cite{Goddard:1983at}. However, we did not figure out how to exploit
this fact before the problem was solved by others.

Before the end of 1984 there were two other major developments. The
first one was the construction of the {\em heterotic string} by
Gross, Harvey, Martinec, and Rohm
\cite{Gross:1984dd,Gross:1985fr,Gross:1985rr}. Their construction
actually accommodated both of the gauge groups. The second one was
the demonstration by Candelas, Horowitz, Strominger, and Witten that
{\em Calabi--Yau compactifications} of the $E_8\times E_8$ heterotic
string give supersymmetric four-dimensional effective theories with
many realistic features \cite{Candelas:1985en}.

By the beginning of 1985, superstring theory -- with the goal of
unification -- had become a mainstream activity. In fact, there was
a very sudden transition from benign neglect to unbounded euphoria,
both of which seemed to me to be unwarranted. After a while, most
string theorists developed a more realistic assessment of the
problems and challenges that remained.

\section{Postscript}

The construction of a dual string theory description of QCD is still
an actively pursued goal. It now appears likely that every
well-defined (finite or asymptotically free) four-dimensional gauge
theory has a string theory dual in a curved background geometry with
five noncompact dimensions. The extra dimension corresponds to the
energy scale of the gauge theory. The cleanest and best understood
example of such a duality is the correspondence between ${\cal N}=4$
supersymmetric Yang-Mills theory with an $SU(N)$ gauge group and
type IIB superstring theory in an $AdS_5 \times S^5$ spacetime with
$N$ units of five-form flux threading the sphere
\cite{Maldacena:1997re}. In particular, the (off-shell)
energy--momentum tensor of the four-dimensional gauge theory
corresponds to the (on-shell) graviton in five dimensions.

Such possibilities were not contemplated in the early years, so it
understandable that success was not achieved. Moreover, the dual
description of QCD is likely to be considerably more complicated
than the example described above. For one thing, for realistic
numbers of colors and flavors, the five-dimensional geometry is
expected to have string-scale curvature, so that a supergravity
approximation will not be helpful. However, it might still be
possible to treat the inverse of the number of colors as small, so
that a semiclassical string theory approximation (corresponding to
the planar approximation to the gauge theory) can be used. If one is
willing to sacrifice quantitative precision, one can already give
constructions that have the correct qualitative features of QCD. One
of their typical unrealistic features is that the Kaluza--Klein
scale is comparable to the QCD scale. I remain optimistic that a
correct construction of a string theory configuration that is dual
to QCD exists. However, finding it and analyzing it might take a
long time.

\section*{Acknowledgments}

I am grateful to Lars Brink for reading the manuscript and making
several helpful suggestions. I also wish to acknowledge the
hospitality of the Galileo Galilei Institute and the Aspen Center
for Physics. This work was partially supported in part by the U.S.
Dept. of Energy under Grant No. DE-FG03-92-ER40701.



\begin{thebibliography}{99}

\bibitem{Schwarz:2000yx}
J.~H.~Schwarz, ``Reminiscences of Collaborations with Jo\"el
Scherk,'' arXiv:hep-th/0007117.

\bibitem{Schwarz:2000yy}
J.~H.~Schwarz, ``String Theory: The Early Years,''
arXiv:hep-th/0007118.

\bibitem{Schwarz:2000dm}
J.~H.~Schwarz, ``String Theory Origins of Supersymmetry,'' \NPPS
{\bf 101}, 54 (2001) [arXiv:hep-th/0011078].

\bibitem{Schwarz:1985ws}
J.~H.~Schwarz, {\it Superstrings -- The First Fifteen Years of
Superstring Theory}, Reprints and Commentary in 2 Volumes, World
Scientific, 1985.

\bibitem{Green:1987cup}
M. B. Green, J.~H.~Schwarz, and E. Witten, {\it Superstring Theory}
in 2 Volumes, Cambridge Univ. Press, 1987.

\bibitem{Becker:2007cup}
K.~Becker, M.~Becker, and J.~H.~Schwarz, {\it String Theory and M-Theory:
A Modern Introduction}, Cambridge Univ. Press, 2007.


\bibitem{Gross:1965}
D. J. Gross and J. H. Schwarz, ``Normal-Threshold Sheet Structure of
Two-Particle Scattering Amplitudes" \PR {\bf 140}, B1054 (1965).

\bibitem{Chew:1961a}
G.~F.~Chew, {\it The S-Matrix Theory of Strong Interactions}, W. A.
Benjamin and Co., 1961.

\bibitem{Chew:1966a}
G.~F.~Chew, {\it The Analytic S-Matrix: A Basis for Nuclear
Democracy}, W. A. Benjamin and Co., 1966.

\bibitem{Chew:1961yz}
G.~F.~Chew and S.~C.~Frautschi, ``Principle of Equivalence for All
Strongly Interacting Particles Within the S Matrix Framework,'' \PRL
{\bf 7}, 394 (1961).

\bibitem{Chew:1962eu}
G.~F.~Chew and S.~C.~Frautschi, ``Regge Trajectories and the
Principle of Maximum Strength for Strong Interactions,'' \PRL {\bf
8}, 41 (1962).

\bibitem{Dolen:1967}
R. Dolen, D.~Horn, and C.~Schmid, ``Prediction of Regge Parameters
of $\rho$ Poles from Low-Energy $\pi N$ Data,'' \PRL {\bf 19}, 402
(1967).

\bibitem{Igi:1967a}
K.~Igi and S.~Matsuda, ``New Sum Rules and Singularities in the
Complex J Plane," \PRL {\bf 18}, 625 (1967).

\bibitem{Igi:1967b}
K.~Igi and S.~Matsuda, ``Some Consequences from Superconvergence for
$\pi N$ Scattering," \PR {\bf 163}, 1622 (1967).

\bibitem{Logunov:1967}
A.~Logunov, L.~D.~Soloviev, and A.~N.~Tavkhelidze, ``Dispersion Sum
Rules and High Energy Scattering,'' \PL {\bf 24B}, 181 (1967).

\bibitem{Dolen:1968}
R. Dolen, D.~Horn, and C.~Schmid, ``Finite-Energy Sum Rules and
Their Application to $\pi N$ Charge Exchange,'' \PR {\bf 166}, 1768
(1968).

\bibitem{Freund:1968}
P.~G.~O.~Freund, ``Finite Energy Sum Rules and Bootstraps,'' \PRL
{\bf 20}, 235 (1968).

\bibitem{Harari:1969}
H.~Harari, ``Duality Diagrams,'' \PRL  {\bf 22}, 562 (1969).

\bibitem{Rosner:1969}
J.~L.~Rosner, ``Graphical Form of Duality,'' \PRL {\bf 22}, 689
(1969).

\bibitem{Veneziano:1968}
G.~Veneziano, ``Construction of a Crossing-Symmetric Regge-Behaved
Amplitude for Linearly Rising Regge Trajectories,'' \NC {\bf 57A},
190 (1968).

\bibitem{Lovelace:1968}
C.~Lovelace, ``A Novel Application of Regge Trajectories,'' \PL {\bf
28B}, 264 (1968).

\bibitem{Shapiro:1969}
J.~A.~Shapiro, ``Narrow Resonance Model with Regge Behavior for $\pi
\pi$ scattering,'' \PR {\bf 179}, 1345 (1969).

\bibitem{Paton:1969}
J.~E.~Paton and H.~Chan, ``Generalized Veneziano Model with
Isospin,'' \NP {\bf B10}, 516 (1969).

\bibitem{Virasoro:1969a}
M.~Virasoro, ``Alternative Constructions of Crossing-Symmetric
Amplitudes with Regge Behavior," \PR {\bf 177}, 2309 (1969).

\bibitem{Bardakci:1969a}
K.~Bardakci and H.~Ruegg, ``Reggeized Resonance Model for Arbitrary
Production Processes,'' \PR {\bf 181}, 1884 (1969).

\bibitem{Goebel:1969}
C.~J.~Goebel and B.~Sakita, ``Extension of the Veneziano Formula to
$N$-Particle Amplitudes,'' \PRL {\bf 22}, 257 (1969).

\bibitem{Chan:1969b}
H.~M.~Chan and T.~S.~Tsun, ``Explicit Construction of the $N$-Point
Function in the Generalized Veneziano Model,'' \PL {\bf 28B}, 485
(1969).

\bibitem{Koba:1969a}
Z.~Koba and H.~B.~Nielsen, ``Reaction Amplitudes for $N$ Mesons, a
Generalization of the Veneziano--Bardakci--Ruegg--Virasoro Model,''
\NP {\bf B10}, 633 (1969).

\bibitem{Koba:1969b}
Z.~Koba and H.~B.~Nielsen, ``Manifestly Crossing-Invariant
Parametrization of the $N$-Meson Amplitude,'' \NP {\bf B12}, 517
(1969).

\bibitem{Shapiro:1970}
J.~A.~Shapiro, ``Electrostatic Analog for the Virasoro Model,'' \PL
{\bf 33B}, 361 (1970).

\bibitem{Fubini:1969a}
S.~Fubini and G.~Veneziano, ``Level Structure of Dual Resonance
Models,'' \NC {\bf 64A}, 811 (1969).

\bibitem{Fubini:1969b}
S.~Fubini, D.~Gordon, and G.~Veneziano, ``A General Treatment of
Factorization in Dual Resonance Models,'' \PL {\bf 29B}, 679 (1969).

\bibitem{Bardakci:1969b}
K.~Bardakci and S.~Mandelstam, ``Analytic Solution of the
Linear-Trajectory Bootstrap," \PR {\bf 184}, 1640 (1969).

\bibitem{Fubini:1970}
S.~Fubini and G.~Veneziano, ``Duality in Operator Formalism,'' \NC
{\bf 67A}, 29 (1970).

\bibitem{Nambu:1970a}
Y.~Nambu, ``Quark Model and the Factorization of the Veneziano
Model,'' p. 269 in Proc. Intern. Conf. on Symmetries and Quark
Models, Wayne State Univ., 1969 (Gordon and Breach, NY 1970).
Reprinted in {\it Broken Symmetry: selected papers of Y. Nambu},
eds. T. Eguchi and K. Nishijima, World Scientific (1995).

\bibitem{Nambu:1970b}
Y.~Nambu, ``Duality and Hadrodynamics'' Lectures at the Copenhagen
Summer Symposium (1970). Reprinted in {\it Broken Symmetry: selected
papers of Y. Nambu}, eds. T. Eguchi and K. Nishijima, World
Scientific (1995).

\bibitem{Susskind:1970e}
L.~Susskind, ``Dual-Symmetric Theory of Hadrons I,'' \NC {\bf 69A},
457 (1970).

\bibitem{Susskind:1970f}
G.~Frye, C.~W.~Lee, and L.~Susskind, ``Dual-Symmetric Theory of
Hadrons. II. - Baryons,'' \NC {\bf 69A}, 497 (1970).

\bibitem{Nielsen:1970a}
H.~B.~Nielsen, ``An Almost Physical Interpretation of the $N$-Point
Veneziano Model,'' submitted to Proc. of the XV Int. Conf. on High
Energy Physics (Kiev, 1970), unpublished.

\bibitem{Fairlie:1970}
D.~B.~Fairlie and H.~B.~Nielsen, ``An Analogue Model for KSV
Theory,'' \NP {\bf B20}, 637 (1970).

\bibitem{Gross:1970a}
D.~J.~Gross, A.~Neveu, J.~Scherk, and J.~H.~Schwarz, ``The Primitive
Graphs of Dual Resonance Models,'' \PL {\bf 31B}, 592 (1970).

\bibitem{Kikkawa:1969}
K.~Kikkawa, B.~Sakita and M.~A.~Virasoro, ``Feynman-like Diagrams
Compatible with Duality. I: Planar Diagrams,'' \PR {\bf 184}, 1701
(1969).

\bibitem{Gross:1970b}
D.~J.~Gross, A.~Neveu, J.~Scherk, and J.~H.~Schwarz,
``Renormalization and Unitarity in the Dual Resonance Model,'' \PR
{\bf D2}, 697 (1970).

\bibitem{Frye:1970b}
G.~Frye and L.~Susskind, ``Non-Planar Dual Symmetric Loop Graphs and
the Pomeron,'' \PL {\bf 31B}, 589 (1970).

\bibitem{Lovelace:1971}
C.~Lovelace, ``Pomeron Form Factors and Dual Regge Cuts,'' \PL {\bf
34B}, 500 (1971).

\bibitem{Virasoro:1970}
M.~Virasoro, ``Subsidiary Conditions and Ghosts in Dual Resonance
Models,'' \PR {\bf D1}, 2933 (1970).

\bibitem{Fubini:1971}
S.~Fubini and G.~Veneziano, ``Algebraic Treatment of Subsidiary
Conditions in Dual Resonance Models,'' \AP {\bf 63}, 12 (1971).

\bibitem{Brink:1973qm}
L.~Brink and D.~I.~Olive, ``The Physical State Projection Operator
in Dual Resonance Models for the Critical Dimension of Space-Time,''
\NP {\bf B56}, 253 (1973).

\bibitem{Brink:1973gi}
L.~Brink and D.~I.~Olive, ``Recalculation of the Unitary Single
Planar Dual Loop in the Critical Dimension of Space-Time,'' \NP {\bf
B58}, 237 (1973).

\bibitem{Goddard:1973}
P.~Goddard, J.~Goldstone, C.~Rebbi and C.~B.~Thorn, ``Quantum
Dynamics of a Massless Relativistic String,'' \NP {\bf B56}, 109
(1973).

\bibitem{Mandelstam:1973}
S.~Mandelstam, ``Interacting String Picture of Dual Resonance
Models,'' \NP {\bf B64}, 205 (1973).


\bibitem{Ramond:1971b}
P.~Ramond, ``Dual Theory for Free Fermions,'' \PR {\bf D3}, 2415
(1971).

\bibitem{Neveu:1971b}
A.~Neveu and J.~H.~Schwarz, ``Factorizable Dual Model of Pions,''
\NP {\bf B31}, 86 (1971).

\bibitem{Neveu:1971fz}
A.~Neveu and J.~H.~Schwarz, ``Tachyon-Free Dual Model with a
Positive-Intercept Trajectory,'' \PL {\bf 3B4}, 517 (1971).

\bibitem{Neveu:1971c}
A.~Neveu, J.~H.~Schwarz, and C.~B.~Thorn, ``Reformulation of the
Dual Pion Model,'' \PL {\bf 35B}, 529 (1971).

\bibitem{Neveu:1971d}
A.~Neveu and J.~H.~Schwarz, ``Quark Model of Dual Pions,'' \PR {\bf
D4}, 1109 (1971).

\bibitem{Thorn:1971}
C.~B.~Thorn, ``Embryonic Dual Model for Pions and Fermions,'' \PR
{\bf D4}, 1112 (1971).

\bibitem{Gervais:1971b}
J.~L.~Gervais and B.~Sakita, ``Field Theory Interpretation of
Supergauges in Dual Models,'' \NP {\bf B34}, 632 (1971).

\bibitem{Brink:1976sz}
L.~Brink, S.~Deser, B.~Zumino, P.~Di Vecchia and P.~S.~Howe, ``Local
Supersymmetry for Spinning Particles,'' \PL {\bf B64}, 435 (1976).

\bibitem{Brink:1976}
L. Brink, P.~Di~Vecchia, and P.~Howe, ``A Locally Supersymmetric and
Reparametrization Invariant Action for the Spinning String,'' Phys.
Lett. {\bf 65B}, 471 (1976).

\bibitem{Deser:1976}
S.~Deser and B.~Zumino, ``A Complete Action for the Spinning
String,'' Phys. Lett. {\bf 65B}, 369 (1976).

\bibitem{Zumino:1974}
B.~Zumino, ``Relativistic Strings and Supergauges,'' p. 367 in {\it
Renormalization and Invariance in Quantum Field Theory,} ed. E.
Caianiello (Plenum Press, 1974).

\bibitem{Wess:1974}
J.~Wess and B.~Zumino, ``Supergauge Transformations in Four
Dimensions,'' \NP {\bf B70}, 39 (1974).

\bibitem{Golfand:1971iw}
Yu.~A.~Golfand and E.~P.~Likhtman, ``Extension of the Algebra of
Poincar\'e Group Generators and Violation of P Invariance,'' {\it
JETP Lett.}\  {\bf 13}, 323 (1971) [{\it Pisma Zh.\ Eksp.\ Teor.\
Fiz.}\ {\bf 13}, 452 (1971)].

\bibitem{Schwarz:1972gr}
J.~H.~Schwarz, ``Dual-Pion Model Satisfying Current-Algebra
Constraints,'' \PR {\bf D5}, 886 (1972).

\bibitem{Brower:1972}
R.~C.~Brower, ``Spectrum Generating Algebra and No-Ghost Theorem for
the Dual Model,'' \PR {\bf D6}, 1655 (1972).

\bibitem{DelGiudice:1972}
E.~Del Giudice, P.~Di Vecchia and S.~Fubini, ``General Properties of
the Dual Resonance Model,'' \AP {\bf 70}, 378 (1972).

\bibitem{Schwarz:1972}
J.~H.~Schwarz, ``Physical States and Pomeron Poles in the Dual Pion
Model,'' \NP {\bf B46}, 61 (1972).

\bibitem{Brower:1973}
R.~C.~Brower and K.~A.~Friedman, ``Spectrum Generating Algebra and
No Ghost Theorem for the Neveu--Schwarz Model,'' \PR {\bf D7}, 535
(1973).

\bibitem{Goddard:1972}
P.~Goddard and C.~B.~Thorn, ``Compatibility of the Dual Pomeron with
Unitarity and the Absence of Ghosts in the Dual Resonance Model,''
\PL {\bf 40B}, 235 (1972).

\bibitem{Gervais:1973}
J.~L.~Gervais and B.~Sakita, ``Ghost-free String Picture of
Veneziano model,'' \PRL  {\bf 30}, 716 (1973).

\bibitem{Olive:1973}
D.~Olive and J.~Scherk, ``No-Ghost Theorem for the Pomeron Sector of
the Dual Model,'' \PL {\bf 44B}, 296 (1973).

\bibitem{Corrigan:1974}
E.~F.~Corrigan and P.~Goddard, ``The Absence of Ghosts in the Dual
Fermion Model,'' \NP {\bf B68}, 189 (1974).

\bibitem{Olive:1974sv}
D.~I.~Olive and J.~Scherk, ``Towards Satisfactory Scattering
Amplitudes for Dual Fermions,'' \NP {\bf B64}, 334 (1973).

\bibitem{Wu:1973}
J.~H.~Schwarz and C.~C.~Wu, ``Evaluation of Dual Fermion
Amplitudes,'' \PL {\bf 47B} (1973) 453.

\bibitem{Corrigan:1974vz}
E.~Corrigan, P.~Goddard, R.~A.~Smith and D.~I.~Olive, ``Evaluation
of the Scattering Amplitude for Four Dual Fermions,'' \NP {\bf B67}
(1973) 477.

\bibitem{Schwarz:1974a}
J.~H.~Schwarz, ``Off-Mass-Shell Dual Amplitudes Without Ghosts,''
\NP {\bf B65} (1973) 131.

\bibitem{Schwarz:1974b}
J.~H.~Schwarz and C.~C.~Wu, ``Off-Mass-Shell Dual Amplitudes. 2,''
\NP  B {\bf B72}, 397 (1974).

\bibitem{Schwarz:1974c}
J.~H.~Schwarz, ``Off-Mass-Shell Dual Amplitudes. 3,'' \NP {\bf B76},
93 (1974).

\bibitem{Ademollo:1976pp}
M.~Ademollo {\it et al.}, ``Dual String with U(1) Color Symmetry,''
\NP {\bf B111}, 77 (1976).

\bibitem{Siegel:1992ev}
W.~Siegel, ``The $N=4$ String is the Same as the $N=2$ String,''
\PRL {\bf 69}, 1493 (1992) [arXiv:hep-th/9204005].

\bibitem{Neveu:1971ma}
A.~Neveu and C.~B.~Thorn, ``Chirality in Dual Resonance Models,''
Phys.\ Rev.\ Lett.\  {\bf 27} (1971) 1758.

\bibitem{Schwarz:1972gr}
J.~H.~Schwarz, ``Dual-Pion Model Satisfying Current-Algebra Constraints,''
Phys.\ Rev.\  D {\bf 5} (1972) 886.

\bibitem{Yang:1954ek}
C.~N.~Yang and R.~L.~Mills, ``Conservation of Isotopic Spin and
Isotopic Gauge Invariance,'' Phys.\ Rev.\  {\bf 96} (1954) 191.

\bibitem{Neveu:1972}
A.~Neveu and J.~Scherk, ``Connection Between Yang--Mills Fields and
Dual Models,'' \NP {\bf B36}, 155 (1972).

\bibitem{Scherk:1971xy}
J.~Scherk, ``Zero-Slope Limit of the Dual Resonance Model,'' \NP
{\bf B31}, 222 (1971).

\bibitem{Weinberg:1965rz}
S.~Weinberg, ``Photons and Gravitons in Perturbation Theory:
Derivation of Maxwell's and Einstein's Equations,''
Phys.\ Rev.\  {\bf 138} (1965) B988.

\bibitem{Scherk:1974a}
J.~Scherk and J.~H.~Schwarz, ``Dual Models for Non-Hadrons,'' \NP
{\bf B81}, 118 (1974).

\bibitem{Yoneya:1973}
T.~Yoneya, ``Quantum Gravity and the Zero Slope Limit of the
Generalized Virasoro Model,'' \NCL  {\bf 8}, 951 (1973).

\bibitem{Yoneya:1974}
``Connection of Dual Models to Electrodynamics and Gravidynamics,''
\PTP {\bf 51}, 1907 (1974).

\bibitem{Green:2007zzb}
M.~B.~Green, H.~Ooguri and J.~H.~Schwarz,
``Decoupling Supergravity from the Superstring,''
Phys.\ Rev.\ Lett.\  {\bf 99} (2007) 041601
[arXiv:0704.0777 [hep-th]].

\bibitem{Scherk:1975a}
J.~Scherk and J.~H.~Schwarz, ``Dual Model Approach to a
Renormalizable Theory of Gravitation,'' Submitted to the 1975
Gravitation Essay Contest of the Gravity Research Foundation.
Reprinted in {\it Superstrings, Vol. 1}, ed. J. Schwarz, World
Scientific (1985).

\bibitem{Freedman:1976}
D.~Z.~Freedman, P.~van Nieuwenhuizen and S.~Ferrara, ``Progress
Toward a Theory of Supergravity,'' \PR {\bf D13}, 3214 (1976).

\bibitem{Deser:1976eh}
S.~Deser and B.~Zumino, ``Consistent Supergravity,'' \PL {\bf B62},
335 (1976).

\bibitem{Brink:1976vg}
L.~Brink and J.~H.~Schwarz, ``Local Complex Supersymmetry in Two
Dimensions,'' \NP {\bf B121}, 285 (1977).

\bibitem{Polyakov:1981rd}
A.~M.~Polyakov, ``Quantum Geometry of Bosonic Strings,'' \PL {\bf
B103}, 207 (1981).

\bibitem{Polyakov:1981re}
A.~M.~Polyakov, ``Quantum Geometry of Fermionic Strings,'' \PL {\bf
B103}, 211 (1981).

\bibitem{Gliozzi:1976}
F.~Gliozzi, J.~Scherk, and D.~Olive, `Supergravity and the Spinor
Dual Model,'' \PL {\bf 65B}, 282 (1976).

\bibitem{Gliozzi:1977}
F.~Gliozzi, J.~Scherk, and D.~Olive, ``Supersymmetry, Supergravity
Theories and the Dual Spinor Model,'' \NP {\bf B122}, 253 (1977).

\bibitem{Jacobi:1829}
C.~G.~J. Jacobi, {\it Fundamenta}, Konigsberg, 1829.

\bibitem{Brink:1977}
L. Brink, J. H. Schwarz,  and J. Scherk , ``Supersymmetric
Yang--Mills Theories,'' \NP {\bf B121}, 77 (1977).

\bibitem{Nahm:1977tg}
W.~Nahm, ``Supersymmetries and Their Representations,'' \NP {\bf
B135}, 149 (1978).

\bibitem{Cremmer:1978}
E.~Cremmer, B.~Julia, and J.~Scherk, ``Supergravity Theory in 11
Dimensions,'' \PL {\bf76B}, 409 (1978).

\bibitem{Scherk:1979a}
J.~Scherk and J.~H.~Schwarz, ``Spontaneous Breaking of Supersymmetry
Through Dimensional Reduction,'' \PL {\bf B82}, 60 (1979).

\bibitem{Scherk:1979b}
J.~Scherk and J.~H.~Schwarz, ``How to Get Masses from Extra
Dimensions,'' \NP {\bf B153}, 61 (1979).

\bibitem{Cremmer:1979}
E. Cremmer, J.~Scherk and J.~H.~Schwarz, ``Spontaneously Broken $N =
8$ Supergravity,'' \PL {\bf B84}, 83 (1979).

\bibitem{Townsend:1995kk}
P.~K.~Townsend, ``The Eleven-Dimensional Supermembrane Revisited,''
Phys.\ Lett.\  B {\bf 350} (1995) 184 [arXiv:hep-th/9501068].

\bibitem{Witten:1995ex}
E.~Witten, ``String Theory Dynamics in Various Dimensions,''
Nucl.\ Phys.\  B {\bf 443} (1995) 85 [arXiv:hep-th/9503124].


\bibitem{Green:1980zg}
M.~B.~Green and J.~H.~Schwarz, ``Supersymmetrical Dual String
Theory,'' \NP  {\bf B181}, 502 (1981).

\bibitem{Green:1981xx}
M.~B.~Green and J.~H.~Schwarz, ``Supersymmetrical Dual String
Theory. 2. Vertices And Trees,'' \NP  {\bf B198}, 252 (1982).

\bibitem{Green:1981ya}
M.~B.~Green and J.~H.~Schwarz, ``Supersymmetrical Dual String
Theory. 3. Loops and Renormalization,'' \NP {\bf B198}, 441 (1982).

\bibitem{Green:1981yb}
M.~B.~Green and J.~H.~Schwarz, ``Supersymmetrical String Theories,''
\PL {\bf B109}, 444 (1982).

\bibitem{Green:1982sw}
M.~B.~Green, J.~H.~Schwarz and L.~Brink, ``$N=4$ Yang--Mills and
$N=8$ Supergravity as Limits of String Theories,'' \NP {\bf B198},
474 (1982).

\bibitem{Green:1982tc}
M.~B.~Green and J.~H.~Schwarz, ``Superstring Interactions,'' \NP
{\bf B218}, 43 (1983).

\bibitem{Green:1983hw}
M.~B.~Green, J.~H.~Schwarz and L.~Brink, ``Superfield Theory of Type
II Superstrings,'' \NP {\bf B219}, 437 (1983).

\bibitem{Green:1984fu}
M.~B.~Green and J.~H.~Schwarz, ``Superstring Field Theory,'' \NP
{\bf B243}, 475 (1984).

\bibitem{Brink:1981nb}
L.~Brink and J.~H.~Schwarz, ``Quantum Superspace,'' \PL {\bf B100},
310 (1981).

\bibitem{Casalbuoni:1976tz}
R.~Casalbuoni, ``The Classical Mechanics for Bose-Fermi Systems,''
\NC {\bf A33}, 389 (1976).

\bibitem{Green:1983wt}
M.~B.~Green and J.~H.~Schwarz, ``Covariant Description of
Superstrings,'' \PL\ {\bf B136}, 367 (1984).

\bibitem{Green:1983sg}
M.~B.~Green and J.~H.~Schwarz, ``Properties of the Covariant
Formulation of Superstring Theories,'' \NP {\bf B243}, 285 (1984).

\bibitem{Green:1982tk}
M.~B.~Green and J.~H.~Schwarz, ``Extended Supergravity in Ten
Dimensions,'' \PL {\bf B122}, 143 (1983).

\bibitem{Schwarz:1983wa}
J.~H.~Schwarz and P.~C.~West, ``Symmetries and Transformations of
Chiral $N=2$, $D=10$ Supergravity,'' \PL {\bf B126}, 301 (1983).

\bibitem{Schwarz:1983qr}
J.~H.~Schwarz, ``Covariant Field Equations of Chiral $N=2$, $D=10$
Supergravity,'' \NP {\bf B226}, 269 (1983).

\bibitem{Howe:1983sr}
P.~S.~Howe and P.~C.~West, ``The Complete $N=2$, $D=10$
Supergravity,'' \NP {\bf B238}, 181 (1984).

\bibitem{Schwarz:1982md}
J.~H.~Schwarz, ``Gauge Groups for Type I Superstrings,'' p. 233 in
Proc. Johns Hopkins Workshop (1982).

\bibitem{Marcus:1982fr}
N.~Marcus and A.~Sagnotti, ``Tree Level Constraints on Gauge Groups
for Type I Superstrings,'' \PL {\bf B119}, 97 (1982).

\bibitem{AlvarezGaume:1983ig}
L.~Alvarez-Gaume and E.~Witten, ``Gravitational Anomalies,'' \NP
{\bf B234}, 269 (1984).

\bibitem{Green:1984sg}
M.~B.~Green and J.~H.~Schwarz, ``Anomaly Cancellation in
Supersymmetric $D=10$ Gauge Theory and Superstring Theory,'' \PL
{\bf B149}, 117 (1984).

\bibitem{Green:1984qs}
M.~B.~Green and J.~H.~Schwarz, ``The Hexagon Gauge Anomaly in Type I
Superstring Theory,'' \NP {\bf B255}, 93 (1985).

\bibitem{Green:1984ed}
M.~B.~Green and J.~H.~Schwarz, ``Infinity Cancellations in SO(32)
Superstring Theory,'' \PL {\bf B151}, 21 (1985).

\bibitem{Goddard:1983at}
P.~Goddard and D.~I.~Olive, ``Algebras, Lattices and Strings,''
DAMTP-83/22, p. 19 in Proc. Marstrand Nobel Sympos. (1986).

\bibitem{Gross:1984dd}
D.~J.~Gross, J.~A.~Harvey, E.~J.~Martinec and R.~Rohm, ``The
Heterotic String,'' \PRL  {\bf 54}, 502 (1985).

\bibitem{Gross:1985fr}
D.~J.~Gross, J.~A.~Harvey, E.~J.~Martinec and R.~Rohm, ``Heterotic
String Theory. 1. The Free Heterotic String,'' \NP {\bf B256}, 253
(1985).

\bibitem{Gross:1985rr}
D.~J.~Gross, J.~A.~Harvey, E.~J.~Martinec and R.~Rohm, ``Heterotic
String Theory. 2. The Interacting Heterotic String,'' \NP {\bf
B267}, 75 (1986).

\bibitem{Candelas:1985en}
P.~Candelas, G.~T.~Horowitz, A.~Strominger and E.~Witten, ``Vacuum
Configurations for Superstrings,'' \NP {\bf B258}, 46 (1985).

\bibitem{Maldacena:1997re}
J.~M.~Maldacena, ``The Large $N$ Limit of Superconformal Field
Theories and Supergravity,'' \ATMP  {\bf 2}, 231 (1998)
[arXiv:hep-th/9711200].

\end{thebibliography}
\end{document}